\documentclass[twocolumn,aps,prd,groupedaddress,showpacs]
              {revtex4}%
\usepackage{graphicx}
\usepackage{amsmath}
\usepackage{bm}
\input{epsf}
\usepackage{epsf}

\usepackage{relsize}
\RequirePackage{xspace}
\begin{document}


\title{\mbox{Classifying the behavior of noncanonical quintessence}\\[10pt]
}

\author{Dan Li$^{1,2}$ and Robert J. Scherrer$^{2}$} \affiliation{ {\footnotesize $^1$School of Science, Zhejiang University of Science and Technology(ZUST), 318 LiuHe Road, HangZhou, China, 310023\\
$^2$Department of Physics and Astronomy, Vanderbilt University, Nashville, TN 37235}\\}




\begin{abstract}
We derive general conditions for the existence of stable scaling solutions for the evolution of noncanonical quintessence,
with a Lagrangian of the form $\mathcal{L}(X,\phi)=X^{\alpha}-V(\phi)$, for
power-law and exponential potentials when the expansion is dominated by a background
barotropic fluid.  Our results suggest that in most cases, noncanonical quintessence with such potentials does not yield
interesting models for the observed dark energy. When the scaling solution is not an attractor, there is a wide range of
model parameters for which the evolution asymptotically resembles a zero-potential solution with
equation of state parameter $w = 1/(2\alpha -1)$, and oscillatory solutions are also possible for positive
power-law potentials; we derive the conditions on the model parameters which produce
both types of behavior. We investigate thawing
noncanonical models with a nearly-flat potential and derive approximate expressions for the evolution of $w(a)$.
These forms for $w(a)$
differ in a characteristic way from the corresponding expressions for canonical quintessence.
\end{abstract}

\maketitle

\section{Introduction}
Observational evidence
\cite{union08,hicken,Amanullah,Union2,Hinshaw,Ade,Betoule}
indicates that approximately
70\% of the energy density in the
universe is in the form of a negative-pressure component,
called dark energy, with the remaining 30\% in the form of nonrelativistic matter.
The dark energy component can be characterized by its equation of state parameter, $w$,
defined as the ratio of the dark energy pressure to its density:
\begin{equation}
\label{w}
w=p/\rho,
\end{equation}
where a cosmological constant, $\Lambda$, corresponds to $w = -1$ and $\rho = constant$.

While a model with a cosmological constant and cold dark matter ($\Lambda$CDM) is consistent
with current observations,
there are many models for dark energy that predict a dynamical equation
of state.  The most widely studied of these are
quintessence models, in which the dark energy
arises from a time-dependent scalar field, $\phi$
\cite{Wetterich,RatraPeebles,Ferreira,CLW,CaldwellDaveSteinhardt,Liddle,SteinhardtWangZlatev}.
(See Ref. \cite{Copeland} for a review).

In this paper, we examine quintessence models with the noncanonical Lagrangian of the form
\begin{eqnarray}
\label{Lagrangian}
\mathcal{L}(X,\phi)=X^{\alpha}-V(\phi),
\end{eqnarray}
where $X \equiv \dot\phi^2/2$, and
we take $\hbar = c = 8 \pi G = 1$ throughout.
This model has been examined both as a model for inflation \cite{UST,RKK} and
for dark energy \cite{FLH,Unnikrishnan,Das,SahniSen,OGSS}. Refs. \cite{SahniSen,OGSS} are the
previous papers most closely related to our discussion.  In Ref. \cite{SahniSen}, Sahni and Sen
examined a model described by Eq. (\ref{Lagrangian}) for which $V(\phi)$ is a constant, and
they argued that for $\alpha \gg 1$, the scalar field can account for both the dark matter
and dark energy.  In Ref. \cite{OGSS}, Ossoulian et al. examined power-law models
for $V(\phi)$ and provided cases for which the resulting evolution is stable.
We extend and generalize this previous work by deriving the
exact conditions for stability for power law potentials, extending this discussion to exponential
potentials, and examining noncanonical quintessence in the limit where the potential is nearly
flat and the model is close to $\Lambda$CDM.

Canonical quintessence models with power-law or exponential
potentials were among the first to be investigated
\cite{Wetterich,RatraPeebles}, but these models have serious problems in
fitting current observations.  For instance, for a potential
of the form $V(\phi) \propto \phi^{-n}$, the data require $n < 1$
\cite{Ratra1,Ratra2}, which is not particularly natural.  Exponential
potentials fare even worse.  They can lead to ``tracker" solutions,
in which $w$ for the scalar field tracks the the same value as $w$
for the background fluid \cite{RatraPeebles,Ferreira,CLW}. While such
models are interesting, they are not consistent with observations.  On the other hand, power-law and
exponential potentials arise naturally in the context of many different models,
so it is important to see whether the noncanonical models examined
here can resurrect them as viable quintessence models.

We also investigate the behavior of noncanonical models when the potential
is nearly flat and the universe contains both quintessence and nonrelativistic matter.  Models of this kind,
in which the scalar field is initially at rest (``thawing" models \cite{CL}),
are a natural way to produce $w$ near $-1$, in
agreement with observations.  Scherrer and Sen \cite{SS} showed that such
models converge toward a similar evolution at late times independent of
the details of the potential; it is interesting to determine whether this
is also the case for noncanonical quintessence.

In the next section, we rederive solutions for the power-law potential with an expansion dominated
by a background barotropic fluid, and we determine the general conditions for stability.  We then perform
the same calculation for the exponential potential.  In Sec. III, we derive an approximation
for the equation of state when the potential is nearly flat, and we discuss the cosmological implications
of all of our results in Sec. IV.

\section{Background-dominated Evolution}

\subsection{Basic Formalism}

We work throughout with the Lagrangian given by Eq. (\ref{Lagrangian}).
The sound speed in this model is \cite{Unnikrishnan}
\begin{equation}
c_s^2 = \frac{1}{2 \alpha - 1}.
\end{equation}
To ensure that $0 \le c_s^2 \le 1$, we will always assume that $\alpha \ge 1$.
(For a more detailed discussion of this point, see Ref. \cite{FGUW}).
The energy density and pressure for $\phi$ are given by:
\begin{eqnarray}
\label{density}
\rho_{\phi}&=&(2\alpha-1) X^{\alpha}+V(\phi),\\
\label{pressure}
p_{\phi}&=&X^{\alpha}-V(\phi).
\end{eqnarray}

In a spatially-flat Friedmann-Robertson-Walker Universe the equation of motion for $\phi$ is
\begin{eqnarray}
\label{EOM}
(\dot{\phi}^{2\alpha-1})^{\cdot}+3H\dot{\phi}^{2\alpha-1}+\frac{2^{\alpha-1}}{\alpha}\frac{dV}{d\phi}=0,
\end{eqnarray}
which can also be written as
\begin{equation}
\ddot \phi + \frac{3 H \dot \phi}{2 \alpha -1} + 
\frac{V^\prime(\phi)}{\alpha(2\alpha-1)}\left(\frac{2}{\dot\phi^2}\right)^{\alpha-1}=0.
\end{equation}

Consider a universe dominated by a background barotropic fluid such as nonrelativistic matter or energy, characterized
by an equation of state parameter $w_B$ (where, e.g., $w_B=0$ for nonrelativistic matter and $w_B = 1/3$ for radiation).  In this case,
the background density scales as
\begin{equation}
\rho \propto a^{-m},
\end{equation}
where
\begin{equation}
m = 3(1+w_B).
\end{equation}
To simplify our expressions, we will work with $m$ instead of $w_B$.  Then $m=3$ for the matter-dominated epoch
and $m=4$ when the universe is radiation-dominated.  The Hubble parameter is then given by
\begin{equation}
H = \frac{2}{mt}.
\end{equation}
In the derivations that follow, we always take $0 \le m \le 6$, corresponding to $-1 \le w_B \le 1$.

\subsection{Power Law Potential}

Consider first the power-law potential
\begin{eqnarray}
V(\phi)=V_0\phi^n,
\end{eqnarray}
where $n$ can be positive or negative. Then Eq. (\ref{EOM}) becomes
\begin{eqnarray}
\label{EOM1}
(\dot{\phi}^{2\alpha-1})^{\cdot}+\frac{6}{m
t}\dot{\phi}^{2\alpha-1}+\frac{2^{\alpha-1} n V_0}{\alpha}\phi^{n-1}=0.
\end{eqnarray}

Assuming a solution of the form
\begin{equation}
\label{phiscaling}
\phi=C t^{\gamma},
\end{equation}
the coefficient and power index are found to be:
\begin{eqnarray}
\label{scaling1}
\gamma&=&\frac{2\alpha}{2\alpha-n},\nonumber\\
C^{n-2\alpha}&=&-\left[(2\alpha-1)+(2\alpha - n)\frac{6}{mn}\right]\left(\frac{\gamma^{2\alpha}}{V_0 2^\alpha}\right).
\end{eqnarray}

These results are a rederivation of those in Ref. \cite{OGSS}.
Note, however, that in order for the solution to be well-defined, the right-hand side of Eq. (\ref{scaling1}) must be positive.  The second
factor is manifestly positive, which means that
\begin{equation}
(2\alpha-1)+(2\alpha - n)\frac{6}{mn} < 0.
\end{equation}
The conditions for this inequality to be satisfied depend on the value of $(2\alpha -1)m$.
For $(2\alpha-1)m < 6$ (which includes the canonical case), we have
\begin{equation}
\label{tracker1}
n < 0 ~~{\rm or} ~~ n > \frac{12 \alpha}{6 - (2\alpha -1)m},
\end{equation}
while for $(2\alpha -1)m > 6$, the condition for the scaling solution to exist is
\begin{equation}
\label{tracker2}
- \frac{12 \alpha}{(2\alpha -1)m - 6} < n < 0.
\end{equation}

Substituting the above scaling solution into Eqs. (\ref{density})-(\ref{pressure}),
we obtain the equation of state parameter
\begin{equation}
\label{wpower}
1+w= - \frac{\alpha m n}{3(2\alpha-n)}.
\end{equation}
which reduces to the canonical result for $\alpha=1$.

Now we must determine the parameter ranges over which our solutions represent stable attractors.
We will follow the methods used previously in Refs. \cite{RatraPeebles,Liddle}.  
We define the variables
\begin{eqnarray}
x_1&=&\phi,\nonumber\\
\label{xdef}
x_2&=&\dot{\phi}^{2\alpha-1},
\end{eqnarray}
and Eq. (\ref{EOM1}) is transformed into the following form:
\begin{eqnarray}
\dot{x}_1&=&x_2^{1/(2\alpha-1)}\nonumber\\
\dot{x}_2&=&-\frac{6}{m t}x_2-\frac{n2^{\alpha-1}V_0}{\alpha}x_1^{n-1}.
\end{eqnarray}
Making the change of variables
\begin{eqnarray}
u_1&=&\frac{x_1}{x_e}-1,\\
u_2&=&\frac{x_2}{\dot{x}_e^{2\alpha-1}}-1,\\
t &=& e^{\tau},
\end{eqnarray}
where $x_e=Ct^{\gamma}$ is the exact solution, we obtain:
\begin{eqnarray}
u_1^{\prime}&=&-\gamma u_1+\gamma {(1+u_2)^{1/(2\alpha-1)}-\gamma},\nonumber\\
\label{u1u2}
u_2^{\prime}&=&-Bu_2+B{(1+u_1)^{n-1}-B},
\end{eqnarray}
where $B=6/m+(\gamma-1)(2\alpha-1)$, and the prime denotes the derivative with respect to $\tau$.
We linearize Eqs. (\ref{u1u2}) about the
the critical point at $u_1=u_2=0$ and solve for the eigenvalues $\Delta{\pm}$ of small perturbations
about this point:
\begin{eqnarray}
\Delta_{\pm}=\frac{1}{2}\left[-(\gamma+B)\pm\sqrt{(\gamma+B)^2+4\gamma B \frac{n-2\alpha}{2\alpha-1}}\right].
\end{eqnarray}
Stability then requires that the real part of both eigenvalues should be negative.
Ref. \cite{OGSS}
previously derived these eigenvalues, but here we determine the exact conditions on $\alpha$, $m$, and $n$
for which the solutions are stable.

The stability condition is divided into two cases, depending on the values of $m$ and $\alpha$.
Taking $\alpha \ge 1$ and $0 \le m \le 6$, we find,
after some tedious algebra, the following stability conditions:

For $(2\alpha -1)m < 6$:
\begin{equation}
\label{powerstable1}
n < 2\alpha ~~~{\rm or}~~~ n > \frac{2 \alpha (m+6)}{6-(2\alpha - 1)m}.
\end{equation}

For $(2\alpha-1)m > 6$:
\begin{equation}
\label{powerstable2}
-\frac{12 \alpha}{(2\alpha - 1) m - 6} < n < 2 \alpha.
\end{equation}

We must now combine these results with the existence conditions given in Eqs. (\ref{tracker1}) and (\ref{tracker2}) to
derive the final conditions on $\alpha$, $m$, and $n$ that yield stable scaling solutions.  Our final result
is the following set of conditions:

For $(2\alpha -1)m < 6$:
\begin{equation}
\label{final1}
n < 0 ~~~{\rm or}~~~ n > \frac{2 \alpha (m+6)}{6-(2\alpha - 1)m}.
\end{equation}

For $(2\alpha-1)m > 6$:
\begin{equation}
\label{final2}
-\frac{12 \alpha}{(2\alpha - 1) m - 6} < n < 0.
\end{equation}
Eq. (\ref{final1}) reduces to the results of Ref. \cite{Liddle} for the case of $\alpha = 1$.

When the lower bound in Eq. (\ref{final2}) is violated, we find a new set
of late-time attractor solutions.  These are given by
\begin{equation}
\label{flat1}
\dot \phi^{2\alpha -1} \propto t^{-6/m}
\end{equation}
with
\begin{equation}
\label{flat2}
w = \frac{1}{2\alpha -1}.
\end{equation}
It is easy to see that the expression for $\phi$ corresponding to Eq. (\ref{flat1})
is a solution of Eq. (\ref{EOM1}) in the limit where the third term in Eq. (\ref{EOM1})
is negligible (i.e., the potential is nearly constant).  However, when
the lower bound in Eq. (\ref{final2}) is violated, this third
term decays away more rapidly with time than the second term, so that
this is, in fact, the correct asymptotic solution.

The solution given by Eqs. (\ref{flat1}) and (\ref{flat2}) is identical
to the constant-potential solution derived by Sahni and Sen \cite{SahniSen}
for the special case where $V(\phi) =0$.
Our results indicate that their model does not necessarily require an exactly
flat potential; it can be achieved by a sufficiently rapidly decaying
potential; i.e., one for which $n < - 12\alpha/[(2\alpha -1)m-6]$ (although
the model of Ref. \cite{SahniSen} in this case would also require the addition
of a constant to this potential).

There is also another mode of evolution when the solution is no longer
an attractor: for $n >0$ and $n$ even, the potential can support oscillatory
solutions.  These occur more generally for arbitrary $n$ and potentials
of the form 
\begin{equation}
\label{Vosc}
V(\phi) = V_0 |\phi|^n.
\end{equation}
Oscillating canonical scalar fields were first investigated by Turner \cite{Turner},
and later reexamined by many others (see Ref. \cite{Dutta} and references therein).
Unnikrishnan et al. \cite{UST} examined oscillating noncanonical scalar fields and showed that
the period-averaged equation of state parameter is given by:
\begin{eqnarray}
\label{oscilEOS}
<w>=\frac{n-2\alpha}{n(2\alpha-1)+2\alpha}.
\end{eqnarray}

Implicit in Eq. (\ref{oscilEOS}) is the assumption that the oscillation frequency $\nu$ is much greater than the Hubble expansion
rate $H$.  As noted in Ref. \cite{Dutta} for canonical quintessence, as long as $\nu/H$ is an increasing function 
of time, the oscillating solution will be the late-time attractor.  Conversely, if $\nu/H$ decreases with time,
then our power-law solution, Eq. (\ref{phiscaling}) is the late-time attractor, and $\phi$ goes smoothly to zero.

Consider a noncanonical scalar field oscillating in the potential given by Eq. (\ref{Vosc}).  Following Ref. \cite{Dutta},
we note that
it oscillates
between the values $-\phi_{max}$ and $\phi_{max}$, with oscillation frequency
\begin{eqnarray}
\nu=\left(\int_{-\phi_{max}}^{\phi_{max}}\frac{d\phi}{C[\rho_{\phi}-V(\phi)]^{1/{2\alpha}}}\right) ^{-1},
\end{eqnarray}
where
\begin{equation}
C=(2^{\alpha} /(2\alpha-1))^{1/{2\alpha}}. 
\end{equation}
The Hubble expansion rate is simply $H = \sqrt{\rho_T/3}$, where $\rho_T$ is the total energy density in the universe.
For the power-law potential in Eq. (\ref{Vosc}), our expression for $\nu$ can be integrated exactly, and we get
\begin{eqnarray}
\label{oscilcond}
\nu/H=\frac{\sqrt{3}C}{4}\frac{\Gamma(1+\frac{1}{n}-\frac{1}{2\alpha})}{\Gamma(1+\frac{1}{n})\Gamma(1-\frac{1}{2\alpha})}\sqrt{\frac{\rho_{\phi}^{1/\alpha}}{\rho_T}}\frac{1}{\phi_{max}}
\end{eqnarray}

Now we are interested in how $\rho_\phi^{1/2\alpha} \rho_T^{-1/2} \phi_{max}^{-1}$ depends on the scale factor.  When
the universe is dominated by a background fluid, $\rho_T = \rho_B \propto a^{-m}$.  The quantity $\phi_{max}$ is
determined by the requirement that $\rho_\phi = V(\phi_{max})$, and the dependence of $\rho_\phi$ on the scale factor $a$ is determined by
the equation of state parameter in Eq. (\ref{oscilEOS}).  Putting all of these together, we obtain
\begin{eqnarray}
\nu/H &\propto& a^\beta,\\
\beta &=& \frac{[(2\alpha-1)m-6]n+2\alpha (m+6)}{2[(2\alpha-1)n+2\alpha]}.
\end{eqnarray}

Then the condition for oscillatory behavior to be a late-time attractor
is that $\nu/H$ increases with the scale factor, or equivalently, $\beta > 0$.  We then have the
following conditions for oscillatory behavior:

For $(2\alpha -1)m < 6$,
\begin{equation}
\label{osc1}
n < \frac{2\alpha(m+6)}{6 - (2\alpha -1)m}.
\end{equation}

For $(2\alpha -1)m > 6$,
\begin{equation}
\label{osc2}
n > \frac{-2\alpha(m+6)}{(2\alpha -1)m-6}.
\end{equation}

For $(2\alpha -1)m < 6$, Eq. (\ref{osc1}) shows that whenever $n$ lies outside of the bounds
given by Eq. (\ref{final1}),
the late-time evolution corresponds to oscillatory
behavior.  For $(2\alpha -1)m> 6$, we have already discussed what happens when the lower bound in Eq. (\ref{final2}) is violated.
When the upper bound is violated, Eq. (\ref{osc2}) is automatically satisfied, and we again have oscillatory behavior.

\subsection{Exponential Potential}

Now consider the exponential potential:
\begin{eqnarray}
V(\phi)=V_0\exp ( -\lambda \phi),
\end{eqnarray}
with $\lambda > 0$.
For the background fluid dominated epoch, Eq. (\ref{EOM}) is:
\begin{eqnarray}
\label{EOM2}
(\dot{\phi}^{2\alpha-1})^{\cdot}+\frac{6}{m
t}\dot{\phi}^{2\alpha-1}-\frac{2^{\alpha-1}\lambda V_0}{\alpha}\exp ( -\lambda
\phi )=0.
\end{eqnarray}
Taking
$\phi=\ln(u)$ and $u=Ct^{\gamma}$,
we immediately obtain the solution
\begin{eqnarray}
\gamma&=&\frac{2\alpha}{\lambda},\\
\label{Cexp}
C^{\lambda}&=&\frac{2^{\alpha} V_0}{\gamma^{2\alpha}(\frac{6}{m}+1-2\alpha)}.
\end{eqnarray}
In order for a solution to exist, the right-hand side of Eq. (\ref{Cexp}) must be positive, which requires
that
\begin{equation}
\label{expexist}
(2\alpha -1)m < 6.
\end{equation}

Then the time evolution of the scalar field is
\begin{eqnarray}
\label{scalsol2}
\phi=\phi_0+\frac{2\alpha}{\lambda}\ln(t/t_0).
\end{eqnarray}
Substituting this scaling solution into Eqs. (\ref{density}) and (\ref{pressure}), we obtain
\begin{eqnarray}
\label{wexp}
1+w=\frac{\alpha m}{3}.
\end{eqnarray}
For the case of canonical quintessence ($\alpha=1$), we have $w=w_B$, so $\rho_{\phi}$ tracks the background fluid density.
However, for $\alpha > 1$, we find $w > w_B$, so that
$\rho_{\phi}$ always decreases more rapidly than the background fluid density.

As in the case of the power-law potential, we make the following change of variables to linearize Eq. (\ref{EOM2}):
\begin{eqnarray}
x_1&=&u_1+x_e,\\
x_2&=&(1+u_2)\dot{x}_e^{2\alpha-1},\\
t&=&e^{\tau},
\end{eqnarray}
where $x_e=\ln (Ct^{\gamma})$ is the exact solution for the exponential potential and
$x_1$, $x_2$ are defined as in Eq. (\ref{xdef}).

We arrive at
\begin{eqnarray}
u_1^{\prime}&=&\frac{2\alpha}{\lambda}((1+u_2)^{1/(2\alpha-1)}-1),\nonumber\\
u_2^{\prime}&=&-B_1 u_2+B_1 (e^{-\lambda u_1}-1),
\end{eqnarray}
where $B_1={6}/{m}+1-2\alpha$. The eigenvalue solution is then:
\begin{eqnarray}
\Delta_{\pm}=\frac{1}{2}\left[-B_1\pm\sqrt{B_1^2-B_1 \frac{8\alpha}{2\alpha-1}}\right].
\end{eqnarray}

Again, requiring the real part of both eigenvalues to be negative to ensure stability,
we get
\begin{eqnarray}
\label{expbound2}
(2\alpha -1)m < 6.
\end{eqnarray}
Note that this is the same condition as in Eq. (\ref{expexist}).  Thus, whenever this condition is satisfied,
our solution is a stable attractor.  There is, however, one caveat.  For
standard quintessence ($\alpha =1$) and sufficiently
small $\lambda$, the stability of the scaling solution breaks down \cite{CLW}.

Now consider what happens when the bound in Eq.
(\ref{expbound2}) is violated.  Once again,
we see that Eq. (\ref{flat1}) is a solution to Eq. (\ref{EOM2})
whenever the first two terms dominate the third term.  But,
in the limit of large $t$, this will always be the case for the exponential potential whenever
$(2\alpha-1)m > 6$.  
Thus, for $(2\alpha-1)m > 6$, we again have $w = 1/(2\alpha -1)$, corresponding to evolution with $V(\phi) = 0$.

\section{Evolution in a nearly-flat potential}
In the previous section we examined the evolution
of a noncanonical scalar field when the universe is dominated by
a background fluid.  However, at late times,
the universe contains a mixture of dark energy and nonrelativitic matter.  In this case, Eq. (\ref{EOM}) cannot, in general,
be solved exactly.  However, in Ref. \cite{SS} it was shown that for a sufficiently flat potential,
i.e., a potential
satisfying the slow-roll conditions,
\begin{equation}
\label{slow1}
\left(\frac{1}{V} \frac{dV}{d\phi}\right)^2 \ll 1,
\end{equation}
and
\begin{equation}
\label{slow2}
\frac{1}{V}\frac{d^2 V}{d\phi^2} \ll 1,
\end{equation}
with $\dot \phi =0$ initially,
the evolution of the scalar field can be well-approximated analytically, yielding a family of solutions
for which $w$ is close to $-1$, consistent with observations.  While
Eqs. (\ref{slow1}) and (\ref{slow2}) are the slow-roll conditions for inflation,
the evolution of the scalar field is quite different from the inflationary case,
since the expansion of the universe in our case is not dominated by the scalar field alone.
Here we extend
the calculation of Ref. \cite{SS} to noncanonical quintessence.  

When Eqs. (\ref{slow1}) and (\ref{slow2}) are satisfied, the scalar field
rolls only a very short distance along the potential, which can then be well-approximated
as a linear potential, $V(\phi) = V_0 - \beta \phi$, where $\beta$ is a constant.
In this case, Eq. (\ref{EOM}) has the exact solution (cf. Ref. \cite{SS})
\begin{equation}
\label{flatexact}
\dot{\phi}^{2\alpha-1} = \beta \int_{a=a_i}^{a_f} \frac{1}{H(a)}\left(\frac{a}{a_f}\right)^3
\frac{da}{a},
\end{equation}
where $H(a)$ is the Hubble parameter appropriate for a universe containing both nonrelativistic matter and dark energy:
\begin{eqnarray}
H  =  \sqrt{(\rho_{M} + \rho_{\phi})/3}.
\end{eqnarray}
In the slow-roll limit, we can make the approximation that $\rho_{\phi}$ is roughly constant and dominated by the scalar field
potential,
$\rho_{\phi} \approx V_0$, while $\rho_{M} = \rho_{M0} a^{-3}$, where $\rho_{M0}$ is the present-day value of the matter density,
and we take the scale factor to be $a=1$ at the present.

With these approximations, Eq. (\ref{flatexact}) can be integrated to give
\begin{eqnarray}
\dot{\phi}^{2\alpha-1} = \frac{2^{\alpha-1} \beta}{\alpha \sqrt{3 V_0}}
\biggl[\sqrt{1+ \left( \Omega_{\phi 0}^{-1} - 1 \right) a^{-3}}\nonumber \\
\label{slowrollphi}
- \left( \Omega_{\phi 0}^{-1}  - 1\right ) a^{-3} \tanh^{-1} \frac{1}{\sqrt{1 + (\Omega_{\phi
0}^{-1}-1)a^{-3}}}\biggr],
\end{eqnarray}
where $\Omega_{\phi 0} = \rho_{\phi 0}/(\rho_{M0} + \rho_{\phi 0})$.

When $w$ is close to $-1$, both the density and pressure of the scalar field are dominated by the potential, and
the equation of state parameter is roughly $1+w \approx 2 \alpha X^\alpha/V_0$. Combining this with Eq.
(\ref{slowrollphi}) and normalizing to the present-day value of $w$, which we denote $w_0$, we obtain
\begin{eqnarray}
\label{w(a)}
1 + w = (1+ w_0)\Biggl[ \sqrt{1 + (\Omega_{\phi 0}^{-1} - 1)a^{-3}}\nonumber\\
- (\Omega_{\phi 0}^{-1} - 1)a^{-3} \tanh^{-1} \frac{1}{\sqrt{1 + (\Omega_{\phi 0}^{-1}-1)a^{-3}}}
\Biggr]^{\left (\frac{2\alpha}{2\alpha - 1} \right)}\nonumber\\
\label{wpred}
\times \left[\frac{1}{\sqrt{\Omega_{\phi 0}}}
- \left(\frac{1}{\Omega_{\phi 0}} - 1 \right)
\tanh^{-1}\sqrt{\Omega_{\phi 0}}\right]^{- \left( \frac{2\alpha}{2\alpha - 1} \right)}.
\end{eqnarray}
Note that for $\alpha = 1$, we regain the corresponding expression in Ref. \cite{SS}.  It is instructive to
compare this prediction for the behavior of $w(a)$ as $\alpha$ is varied; a graph of $w(a)$ is given in Fig. 1
for several values of $\alpha$, where we have fixed $\Omega_{\phi 0} = 0.7$ and $w_0 = -0.9$.
\begin{figure}[t]
\centerline{\epsfxsize=3.7truein\epsfbox{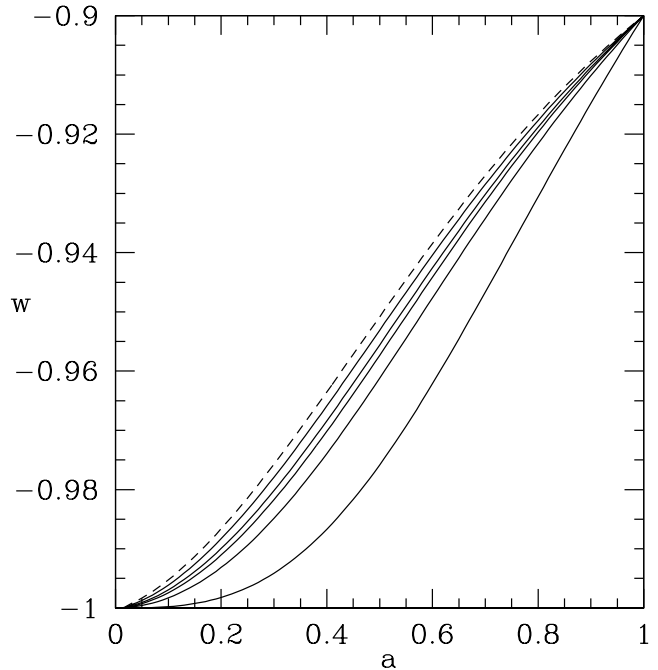}}
\caption{Analytic approximation for the evolution of $w$ as a function of the
scale factor,
$a$, normalized to $a=1$ at the present,
in noncanonical quintessence models with a nearly flat potential,
for $\Omega_{\phi 0} = 0.7$, $w_0 = -0.9$, and (solid curves, bottom to top), $\alpha=1$,
$\alpha=2$, $\alpha=3$, $\alpha=4$, and $\alpha=8$. Dashed curve gives the asymptotic
limit as $\alpha \rightarrow \infty$.}
\end{figure}
Several things are apparent from this figure.  The canonical ($\alpha=1$) case
produces a $w(a)$ that is clearly distinct from the noncanonical cases.  However,
as $\alpha$ increases, the form for $w(a)$ begins to converge to a single behavior independent
of $\alpha$.  This is obvious from Eq. (\ref{w(a)}), since the exponent $2\alpha/(2 \alpha -1)$
goes to $1$ for large $\alpha$.

\section{Discussion}

Our results for noncanonical quintessence with power-law and exponential potentials do
not suggest that such models can provide a better fit to observations than the corresponding canonical
quintessence models; in general,
they will yield a worse fit.  Consider first the power-law case for a
matter-dominated universe.  In that case, Eq. (\ref{wpower}) gives
\begin{equation}
1+w = - \frac{\alpha n}{2\alpha - n}.
\end{equation}
In the canonical ($\alpha=1$) case, the requirement that $w$ not be too far from $-1$ during matter
domination forces $n$
to be negative and close to zero.  However, taking $\alpha > 1$ makes matters
worse, since $n$ must be even closer to zero to obtain a given value
for $w$.  Thus, one gains very little at the expense of the additional
complexity of the model.

On the other hand, moving from canonical to noncanonical quintessence
does produce one interesting new result, which
arises for large
negative potentials, i.e., those which violate the lower bound in Eq.
(\ref{final2}).  In this case, the asymptotic evolution resembles the evolution
in a flat $V(\phi)=0$ potential.  As noted in Ref. \cite{SahniSen}, this does not
provide a dark energy component, but it can mimic, in the limit of large
$\alpha$, dark matter.

The exponential potential, for $\alpha > 1$, does not produce tracking
behavior as it does in canonical quintessence.  Indeed, for a matter-dominated
or radiation-dominated universe ($m=3,4$) and $\alpha \ge 2$, one
again obtains evolution with $w = 1/(2\alpha -1)$, independent of the
parameters of the potential or the background equation of state.
Thus, our results suggest that the behavior outlined in Ref. \cite{SahniSen}
for noncanonical quintessence
is
generic to a wide variety of potentials, not just a flat potential.

Perhaps more interesting is the generic behavior of thawing noncanonical quintessence
in a nearly-flat potential.  In this case $w$ is always close to $-1$, but
the evolution of $w(a)$, for fixed values of $w_0$ and $\Omega_{\phi0}$, is
dependent on the value of $\alpha$, asymptotically approaching a single
functional form in the limit of large $\alpha$.  Thus, there is a useful
signature distinguishing this particular class of noncanonical quintessence models
from the canonical case.

\begin{acknowledgments}
We thank S. Pi, Y. Gao, S. Dutta, H.-Y. Chang, and B. Ratra for helpful discussions and comments.
R.J.S. was supported in part by the Department of Energy (DE-SC0011981).
\end{acknowledgments}

\end{document}